\documentclass[preprint]{elsarticle}

\usepackage{lineno,hyperref}
\modulolinenumbers[5]

\biboptions{authoryear}

\usepackage{amssymb,amsmath}
\usepackage{wasysym} 
\usepackage{color}

\newcommand{\cb}{}

\begin{document}

\begin{frontmatter}

\title{An attempt to detect transient changes in Io's SO$_2$ and NaCl atmosphere}

\author[KTH]{Lorenz Roth}
\ead{lorenz.roth@ee.kth.se}
\author[IRAM]{Jeremie Boissier}
\author[Ames]{Arielle Moullet}
\author[Col1]{\'Alvaro S\'anchez-Monge}
\author[Caltech]{Katherine de Kleer}
\author[Tadano]{Mizuki Yoneda}
\author[UTok]{Reina Hikida}
\author[ISAS]{Hajime Kita}
\author[UToh]{Fuminori Tsuchiya}

\author[KTH]{Aljona Blöcker}
\author[SwRI]{G. Randall Gladstone}
\author[Liege]{Denis Grodent}
\author[KTH]{Nickolay Ivchenko}
\author[LESIA]{Emmanuel Lellouch}
\author[SwRI]{Kurt D. Retherford}
\author[ColG]{Joachim Saur}
\author[Col1]{Peter Schilke}
\author[JHU]{Darrell Strobel}
\author[Col1]{Sven Thorwirth}

\address[KTH]{Space and Plasma Physics, Royal Institute of Technology KTH, Stockholm, Sweden}
\address[IRAM]{IRAM, 300 rue de la Piscine, 38406 St. Martin d’Heres, Grenoble, France} 
\address[Ames]{NASA, Ames Research Center, Space Science Division, Moffett Field, CA, USA}
\address[Col1]{I. Physikalisches Institut, Universität zu Köln, Zülpicher Str. 77, Köln, Germany}
\address[Caltech]{California Institute of Technology, 1200 E California Blvd. M/C 150-21, Pasadena, CA 91125, USA}
\address[Tadano]{Tadano Ltd., Takamatsu, Japan}
\address[UTok]{Department of Earth and Planetary Science, University of Tokyo, Tokyo, Japan}
\address[ISAS]{Institute of Space and Astronautical Science
Japan Aerospace Exploration Agency}
\address[UToh]{Graduate School of Science, Tohoku University, Sendai, Japan}

\address[SwRI]{Southwest Research Institute, San Antonio, TX 78238, USA}
\address[Liege]{Laboratoire de Physique Atmosphérique et Planétaire, STAR Institute, Université de Liège, Liège, Belgium}
\address[LESIA]{LESIA–Observatoire de Paris, CNRS, UPMC Univ. Paris 06, Univ. Denis Diderot, Sorbonne Paris Cite, Meudon, France}
\address[ColG]{Institut für Geophysik und Meteorologie, Universität zu Köln, Albertus-Magnus-Platz, Cologne, D-50923, Germany}
\address[JHU]{Johns Hopkins University, Baltimore, MD, USA}

\begin{abstract}
Io's atmosphere {\cb is predominately SO$_2$} that is sustained by a combination of volcanic outgassing and sublimation. The loss from the atmosphere is the main mass source for Jupiter's large magnetosphere. Numerous previous studies attributed various transient phenomena in Io's environment and Jupiter's magnetosphere to a sudden change in the mass loss from the atmosphere supposedly triggered by a change in volcanic activity. {\cb Since the gas in volcanic plumes does not escape directly}, such causal correlation would require a transient volcano-induced change in atmospheric abundance, which has never been observed so far.

Here we report four observations of atmospheric SO$_2$ and NaCl from the same hemisphere of Io, obtained with the IRAM NOEMA interferometer on 11 December 2016, 14 March, 6 and 29 April 2017. These observations are compared to measurements of volcanic hot spots and Io's neutral and plasma environment. We find a stable NaCl column density in Io's atmosphere on the four dates. The SO$_2$ column density derived for December 2016 is about 30\% lower compared to the SO$_2$ column density found in the period of March to April 2017. This increase  in SO$_2$ from December 2016 to March 2017 might be related to increasing volcanic activity observed at several sites in spring 2017, but the stability of the volcanic trace gas NaCl and resulting decrease in NaCl/SO$_2$ ratio do not support this {\cb interpretation}. Observed dimmings in both the sulfur ion torus and Na neutral cloud suggest rather a decrease in mass loading in the period of increasing SO$_2$ abundance. The dimming Na brightness and stable atmospheric NaCl furthermore dispute an earlier suggested positive correlation of the sodium cloud and the hot spot activity at Loki Patara, which considerably increased in this period. The environment of Io overall appears to be in a quiescent state, preventing further conclusions. Only Jupiter's aurora morphology underwent several short-term changes, which are apparently unrelated to Io's quiescent environment or the relatively stable atmosphere.  

%.... Long-term systematic monitoring of all the involved regions and processes will be required to address this issue.   
\end{abstract}

\begin{keyword}
Io's atmosphere \sep Volcanic activity \sep Magnetospheric variability
\end{keyword}

\end{frontmatter}

%\linenumbers

\section{Introduction}

The atmosphere of Jupiter's volcanic moon Io consists primarily of SO$_2$ and is generated through a combination of direct volcanic outgassing and sublimation of volcanic frost deposits \citep[see e.g. review by][]{lellouch07}. The relative contribution of these two sources is subject of many studies with sometimes ambiguous results \citep[e.g.,][]{clarke94,saurstrobel04,spencer05,retherford07,roth11,tsang15,jessupspencer15,dekleer19-SO,hue19}. Infrared (IR) observations of an SO$_2$ absorption band obtained over a Jupiter season revealed a stable SO$_2$ atmosphere, with moderate but clear dependence on heliocentric distance suggesting that both sublimation and (constant) volcanic outgassing are viable sources \citep{tsang12}. The most compelling evidence for sublimation being the dominating source was finally also provided by IR measurements: \cite{tsang16} detected an SO$_2$ collapse by a factor of 5 $\pm$ 2 after Io entered the shadow of Jupiter, which they explained with a decrease of surface temperature and thus of sublimation. 
 
Besides sulfur and oxygen compounds, NaCl  was detected in Io's atmosphere with submillimeter observations \citep{lellouch03}. The very low NaCl vapor pressure rules out sublimation as effective source to balance the fast loss via photodissociation \citep{moses02-2}. Instead, NaCl must be sustained through volcanic outgassing and {\cb a localized confinement to individual volcanic sites is also consistent with observations} \citep{moullet10,moullet15}. Besides direct outgassing of gaseous NaCl, vaporization of NaCl condensates might become a possible source if volcanic eruptions lead to surface temperatures above 1000 K \citep{vartenberg21} as occasionally observed at Io \citep{mcewen98,keszthelyi07}.

Through various processes, fast and slow Na atoms are generated and ejected from Io \citep{wilson02} forming extended neutral clouds along Io's orbit \citep{brown74,schneider07}. Although only a rare trace species in the magnetosphere, the sodium atoms are easily detected because of their high resonance scattering efficiency and are therefore often used as a diagnostic for monitoring the Jovian neutral environments \citep[e.g.,][]{mendillo90,grava14}.

An average amount of 1 ton/s of material, primarily sulfur and oxygen, is lost from Io's atmosphere to the magnetosphere through collisions with the corotating plasma \citep{broadfoot79}. Model results suggest that 80\% of this mass is lost as neutrals while 20\% is directly ionized in the atmosphere and picked up by the magnetic field \citep{saur03}. Most of the lost neutrals are also eventually ionized and accumulate along Io's orbit where they form the Io plasma torus near the moon as well as the magnetospheric plasma sheet when radially transported outwards \citep[see e.g., review by][]{thomas04}.

The mass loss from Io's atmosphere is the main source of plasma for Jupiter's huge magnetosphere and substantially affects {\cb its} dynamics and extent \citep{krupp04,khurana04}. A striking observable effect of the plasma input and transport within the magnetosphere is Jupiter's bright main auroral emission. The region where the plasma rotation lags Jupiter's corotation magnetically maps to the continuously present main emission of the aurora, initially explained through a stationary field-aligned electric current system \citep{cowleybunce01,hill01}. Recent results from the NASA Juno mission indicate that the generation of Jupiter's main auroral emission is more diverse and dynamic, including broadband or stochastic processes \citep{mauk17,saur18} as well as magnetic loading and unloading \citep{yao19}.   

Hence, it is generally agreed on that the mass loss from Io's atmosphere plays a crucial role for Jupiter's magnetospheric processes. The source of this atmosphere on the other hand is either directly (outgassing) or indirectly (frost deposit sublimation) connected to Io's volcanic activity, which is clearly time-variable \citep[e.g.,][]{williams07}. Given the variable nature of the volcanism and Io's role as primary magnetospheric mass source, various transient changes in the sodium neutral cloud, the plasma and neutral torus, and Jupiter's auroral morphology and brightness were proposed to originate from a sudden change of the atmospheric mass output triggered by changes in volcanic activity such as strong eruptions. We discuss some of these studies in the following.

Earlier studies claiming such volcanic control used data from individual observing campaigns or specific events like the Cassini flyby at the Jupiter system. \cite{brownbouchez97} observed a sharp increase in the sodium cloud followed by an increase in sulfur ion torus emissions and interpreted this as volcanic mass loading event. \cite{delamere04} modeled a change in torus properties detected during the Cassini survey (2000-2001) and derived a decrease of factor 3 within a month in Io's mass loading. \cite{bonfond12} attributed a systematic change of Jupiter’s auroral emission in 2007 to an increase in volcanic activity, based on the imaging of the Tvashtar volcano plume during the flyby of the New Horizons spacecraft \citep{spencer07}. An increase in the sodium brightness happened around the same time \citep{yoneda09}. 

Since 2013, the Io torus emissions as well as the brightness of Jupiter's aurora were monitored systematically during Jupiter observing seasons by the Hisaki space observatory \citep{yoshikawa14}. {\cb The most substantial event documented by Hisaki occurred in February 2015}: The oxygen neutral density in the Io torus increased by factor of $2.5$ \citep{koga18a}, simultaneously with an increase in the sodium cloud brightness seen in ground-based observations \citep{yoneda15}. 
An emission enhancement from singly ionized sulfur in the Io torus was also measured around the same time, and with some delay the emissions from ions of higher charge states increased as well \citep{yoshikawa17,yoshioka18}. Following this period, Hisaki detected more frequent intense auroral brightenings accompanied by changes in the color ratio of auroral UV emission that reflects the energy of precipitating electrons \citep{tsuchiya18,tao18}. All these observations were explained by reconfigurations of the torus (change of densities and temperatures) and magnetosphere after some period of increased mass loading. 

The above mentioned studies claimed that the observed changes were triggered by volcanic activity on Io. The volcanic activity is commonly assessed through observations of thermal IR emissions from volcanic hot spots \citep[e.g.,][]{blaney95,rathbun04}. However, the temporal and causal relation of the magnetospheric events to the thermal hot spots is not clear.

An often cited study by \cite{mendillo04} proposed that the hot spots emission on the sub-Jovian hemisphere, which is dominated by the brightness of Loki Patera, is positively correlated to the Na cloud brightness, but this correlation is derived from a rather small sample size (see their figure 2). {\cb Another method to constrain volcanic outgassing relation is provided by infrared observations of sulfur monoxide (SO) 1.7 $\mu$m forbidden emissions, which are thought to originate directly from outgassed excited SO \citep{depater02,depater07}. Recently, \cite{dekleer19-SO} did not find a significant dependence of the SO emissions to thermal hot spot activity and found a particularly high SO abundance during a time of particularly low thermal emissions at Loki Patera on one occasion. The latter suggests that hot spot activity is not necessarily coupled to plume activity and gas emission.} 

The systematic magnetospheric changes detected by Hisaski in 2015 were {\cb claimed to be} associated with one single strong eruption at Kurdalagon Patera \citep[e.g.,][]{yoshikawa17,tao18,koga18a}. However, the continuous monitoring by \cite{dekleer16a} and \citet{dekleer19-iotime} detected 19 events between 2013 and 2019 that the authors categorized as bright eruptions like the Kurdalagon event in 2015. The argument for Kurdalagon as the trigger of the magnetospheric reconfiguration is only based on the temporal coincidence. It remains unclear why and how the Kurdalagon eruption could have affected the mass loss from Io in a way that a large magnetospheric reconfiguration occurred while other, sometimes brighter, eruptions did not affect the magnetosphere at all. For example, the new bright outburst designated 'UP 254' in May 2018 had an intensity of 125~GW/$\mu$m/sr in the L'-band \citep{dekleer19-iotime} (compared to the maximum of 68~GW/$\mu$m/sr at Kurdalagon) but did not lead to measurable changes in the torus \citep{tsuchiya19-2}. 

{\cb An alternative cause for change in mass loss from Io's, independent of volcanic events, would be an reconfiguration of the Jovian magnetosphere: A change in plasma density and temperature in Io's orbit (through e.g. an interchange event, \cite{bolton97}) affects the plasma-atmosphere interaction and thus the mass loading \citep{saur99}. Such external triggers for transient or intermittent mass loading changes have not been studied so far to our knowledge.}

Some recent studies even provided results that question the basic hypothesis that volcanic outbursts can lead to increased mass loading. In a fortunate IR high-resolution observation, \cite{lellouch15} measured a very strong thermal outburst in the continuum emission over the Pillan Patera. The strong continuum allowed a measurement of atmospheric line absorption in the thermal component from the hot spot, which did not reveal any measurable extra SO$_2$ abundance above this region. This result shows that even extreme thermal eruptions are not necessarily connected to changes in the bulk atmosphere. Furthermore, Monte Carlo simulations by \cite{mcdoniel17} suggest the active Pele plume contributes only about 1\% to the atmospheric content and that the net increase from a plume to the hydrostatic atmosphere density on the dayside is only a fraction of the increase on the night side.

Direct escape of volcanic plume gases is marginal, because the plume ejection velocities are generally below Io's escape velocity of 2.6 km/s. For a ballistic trajectory, the highest observed plumes of 400 km would imply an ejection velocity of 1.2 km/s, the largest estimated gas temperatures of 800 K (for NaCl, \cite{lellouch03}) correspond to a root-mean-square velocity for SO$_2$ of 0.6 km/s.  In addition, simulations revealed that the ejected plume gas is effectively contained by the canopy shocks further reducing the possibility to escape \citep{zhang03,geissler07}. This means that mass loading of volcanic gases {\cb probably} occurs through the same processes as for the global sublimated atmosphere, namely through primarily elastic collisions of ions and neutrals {\cb and secondarily photo-ionization} \citep{saur99,dols12,bloecker18}.

The fact that large amounts of plume gases cannot escape directly means that if volcanic eruptions indeed affect the bulk mass loss temporarily, they also need to lead to measurable short-time changes in the bound atmosphere. However, volcanically induced aperiodic changes in Io's SO$_2$ atmosphere have never been confirmed observationally.  The SO$_2$ survey over 10 years by \cite{tsang12} revealed only seasonal changes but no stronger aperiodic deviations (see e.g. their figure 12).

In this study, we report four submillimeter observations of SO$_2$ and NaCl atmospheric emission lines taken over a period of four months in 2016/2017 with the goal to detect volcanically induced changes. The data acquisition and processing is described in Section \ref{sec:obs}. By fitting results from an atmosphere model to the extracted emission lines, we derive global abundances for SO$_2$ and NaCl on the observed hemisphere (Section \ref{sec:model}). In Section \ref{sec:discuss} , we discuss the obtained abundances on the four different days and in particular the time-variability in the atmosphere and compare them to observations of Io's volcanic hot spots, the sodium cloud, the sulfur ion torus and Jovian aurora from the same period. Section \ref{sec:sum} summarizes the results.

\section{Observations and data processing}
\label{sec:obs}
\subsection{Observations}
Io was observed by the NOrthern Extended Millimetre Array (NOEMA) interferometer of the Institut de Radioastronomie Millimetrique (IRAM) on four occasions, see Table \ref{tab:obs}. The observations were taken in an intermediate array configuration ('C') providing spatial resolution similar to the size of Io's diameter ($\sim$1 arcsec).
A spectral setting was selected such that four SO$_2$ pure rotational emission lines 
($J_{K_a,K_c}=32_{4,28}-32_{3,29}$ at 258.389\,GHz, 
$20_{7,13}-21_{6,16}$ at 258.667\,GHz, 
$9_{3,7}-9_{2,8}$ at 258.942\,GHz,  and $30_{4,26}-30_{3,27}$ 259.599\,GHz) 
were covered simultaneously with a NaCl line ($J=20-19$ at 260.223\,GHz) 
at a resolution of 0.20 MHz (or 0.24~km s$^{-1}$). 

\begin{table}
\caption{Details and geometry parameters of the IRAM/NOEMA observations.}
\begin{tabular}{llllcccc}
\hline
Obs & Observation 	& Start	&   End     &	Sun     	&   Earth		&	Io				& Observed	\\
\#  &  Date 				& time 	&   time   	& 	distance&	distance	&	diameter	&  Io CML$^a$ \\
 &  					& (UT) 	&   (UT)    &	(AU)    	&	(AU)    		&	(arcsec)	&  (degree) \\							
\hline		
%\multicolumn{7}{c}{EXTRA INFO} \\
1   &   2016-Dec-11	& 04:12 &   11:20       & 5.46   &  5.86		&	0.86		& 221 -- 280 \\
2   &   2017-Mar-14	& 22:30 &   03:53$^{+1}$& 5.45   &	4.55		&	1.11		& 219 -- 266 \\
3   &   2017-Apr-06	& 23:10 &   04:04$^{+1}$& 5.45   &	4.45		&	1.13		& 228 -- 269 \\
4   &   2017-Apr-29	& 21:58 &   01:53$^{+1}$& 5.45   &	4.52		&	1.12		& 221 -- 254 \\
\hline
\end{tabular}
\label{tab:obs}
{\cb \footnotesize{$^a$ Io's central meridian (West) longitude as seen from Earth}}
\end{table}

All four tracks were scheduled such that the {\cb anti-Jovian to trailing}  hemisphere is observed, where the atmospheric density is comparably high \citep{spencer05,feaga09,tsang13,jessupspencer15} and a large fraction of the bright transient hot spots is found \citep{dekleer16b}. The similar longitude coverage enables a comparison of the fluxes measured during each of the observations as well as to the volcanic activity of hot spots on the same hemisphere. The ranges in Io's central meridian West longitude for each track (CML in Table \ref{tab:obs}) are roughly centered on 245$^\circ$ W, corresponding to a longitude range of 155 -- 335$^\circ$ W of the entire observed hemisphere from dawn to dusk limb. \\

\subsection{Data processing}

The whole data reduction was performed using the CLIC and MAPPING packages of the GILDAS\footnote{https://www.iram.fr/IRAMFR/GILDAS/} software suite \citep{gildas13}.  For each track, we followed the same sequence of operations consisting of a \textit{standard calibration} (1), \textit{flux calibration} (2), and the \textit{spectral line extraction} (3). The three steps are explained in more detail now, the adopted calibration parameters and references are given in Table \ref{tab:calib}.

\begin{table}
\caption{Data processing and calibration parameters and objects, including the root-mean-square values of the phase ($\phi$ rms) and amplitude (A rms).}
\centering
\begin{tabular}{lcccccccc}
\hline
Obs 			& \multicolumn{3}{c}{RF bandpass calibration} & \multicolumn{3}{c}{Gain calibration} & Flux cal. & Synthesized beam	\\
 \#			& Source &   $\phi$ rms     &	A rms     &   Source & $\phi$ rms & A rms  & Source &  (at 258.942 GHz)	\\
\hline		
%\multicolumn{7}{c}{EXTRA INFO} \\
1	&  3C84  & 0.8--2.4$^\circ$   & 1.3--2.7\%  & 3C279 & 30-90$^\circ$ &17--24\% & Lkh$\alpha$101   &  1.64'' x 0.56''  \\
2	&  0851+202 & 0.5--1.9$^\circ$ &0.5--2.1\% & 3C279 & 15--71$^\circ$& 09--18\% & MWC349           &  2.14'' x 1.23''  \\
3	&  3C273    & 0.9--2.2$^\circ$ &1.2--4.9\% & 3C279 &24--71$^\circ$ & 13--18\% & MWC349           &  2.98'' x 1.74''         \\
4	&  3C279    & 0.6--2.2$^\circ$ &0.8--3.5\% & 3C279 &10--63$^\circ$ & 08--23\% & MWC349           &  3.59'' x 1.47'' \\
\hline
\end{tabular}
\label{tab:calib}
\end{table}

In the first step of the \textit{standard calibration} (1) the system response is calibrated by measuring a continuum on a strong calibrator. In the second step, atmospheric and instrumental phase and amplitude variations are calibrated for, using measurements of a gain calibrator (a point source with phase and amplitude that should remain constant over the observed time). The best phase rms (root-mean-square) after calibration was obtained for 2017-Apr-29 data set while 2016-Dec-11 was observed under poor conditions (Table \ref{tab:calib}). The derived amplitude root-mean-squares (rms) were between $\sim$10 and  $\sim$25\%. In the final step of the standard calibration, the absolute amplitude scale of the data was determined based on observations of one of the secondary flux calibrators for NOEMA where fluxes are monitored against planets all along the year. Given the high frequency and the relatively poor observing conditions this flux calibration provides a global uncertainty on the order of 15\%. 

The \textit{flux calibration} (2) is carried out using Io's continuum emission. Io is commonly used as primary flux calibrator and good model predictions exist for its thermal emission.  Since the goal of our project was to study line emission, we could use Io's thermal emission to improve the flux calibration as described for the last step of the \textit{standard calibration}. In the first part of the flux calibration, the emission from line-free channels in the calibrated visibility spectral tables were averaged to generate continuum visibilities for a first image of Io's continuum emission. Given the high signal to noise ratio of the obtained map, an iteration of self-calibration was then performed to estimate phase gains and apply them to the visibilities, leading to improved image quality. Io's disk integrated flux $S^{\textrm{Io}}_{\textrm{Obs}}$ was obtained by fitting a disk model to the observed visibilities. Finally, a theoretical Io flux $S^{\textrm{Io}}_{\textrm{Model}}$ was extracted from the model available in the Common Astronomy Software Application \citep[CASA,][]{mcmul07}. The model is described in the ALMA memo 594 \citep{butler12}.  The amplitude scale derived in the last step of the standard calibration was then scaled by the factor $S^{\textrm{Io}}_{\textrm{Model}}/S^{\textrm{Io}}_{\textrm{Obs}}$. According to \cite{butler12} the uncertainty on Io's continuum flux should be within 5\% and hence significantly better than the original flux calibration. 

For the \textit{spectral line extraction} (3), continuum visibilities were first obtained again from the line-free channels in the rescaled visibility table. Self-calibration gains were measured again. The success of the flux rescaling was confirmed by repeating this step
and checking that the measured flux of Io matched the value given by the model. The self-calibration gains were then applied to all channels of the visibility table. The continuum emission was subtracted from the visibility spectra by fitting a first order polynomial function to the line-free channels.

\begin{figure}
\center\noindent\includegraphics[width=1.\textwidth]{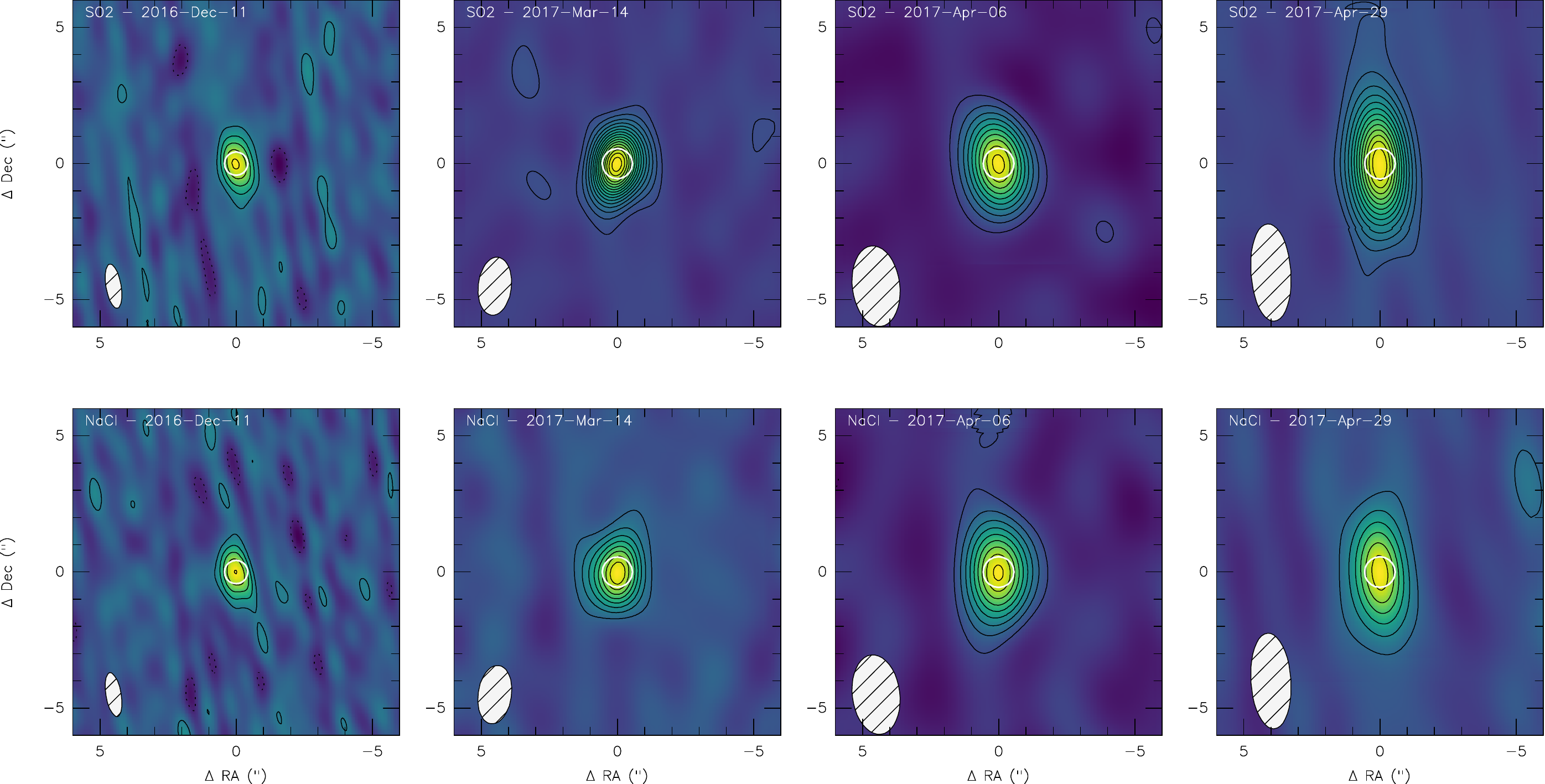}
\caption{IRAM/NOEMA maps of the emission for the 258.942-GHz SO$_2$ line and the NaCl line (260.223 GHz) after self-calibration for the four observation days. The levels of the contour lines (thin black) are in 2-sigma steps. {\cb Io's disk is shown in solid white.} The beam size is illustrated in the inlet in the lower left and given in Table \ref{tab:calib} for the SO$_2$ line. }
\label{fig:IRAMobs}
\end{figure}

\begin{figure}
\center\noindent\includegraphics[width=.95\textwidth]{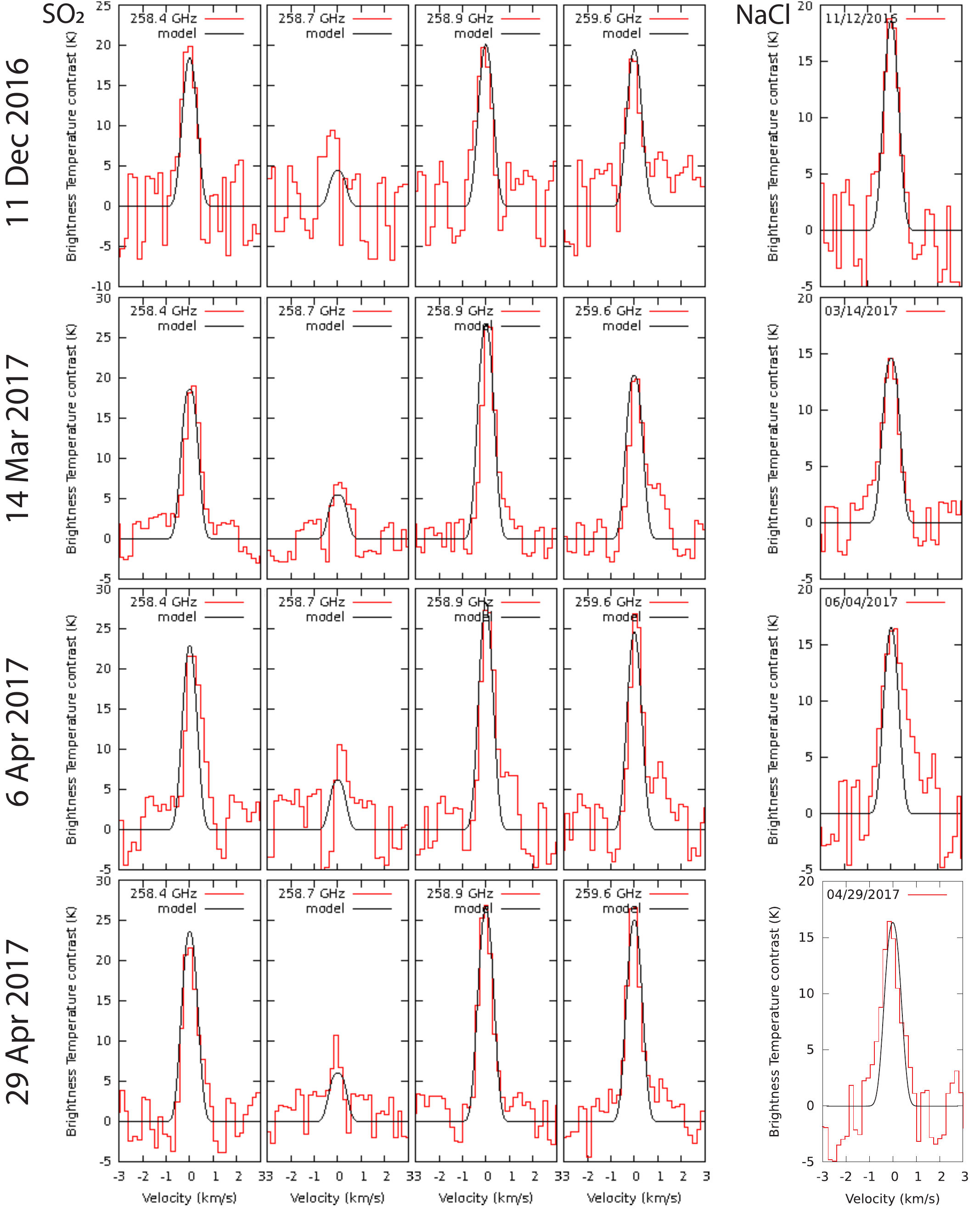}
\caption{Disk-averaged spectra (red) of the four SO$_2$ lines (left four columns) and the NaCl 260.223 GHz line (rightmost column) for the four IRAM observing tracks. The continuum was subtracted. The model results for fitting the brightness temperature (simultaneously for the four SO$_2$ lines) are shown in black. The corresponding images for the SO$_2$ 258.942 GHz line and the NaCl line are shown in Figure \ref{fig:IRAMobs}.}
\label{fig:modelSO2}
\end{figure}

Self-calibrated, line-only visibilities were imaged and deconvolved to build cubes
around all the targeted spectral lines. Line integrated maps of SO$_2$ and NaCl are presented in Figure \ref{fig:IRAMobs}. For each channel, an Io-sized disk model was fitted to the visibilities and the total fluxes  (i.e. fluxes at zero radius in the aperture plane) were used to build the surface integrated spectra, shown in Figure \ref{fig:modelSO2}. 

In order to roughly estimate the fluxes in each line, we fitted Gaussian profiles to the extracted line spectra. The areas under the Gaussian profiles calculated with the fitted peak and width parameters are summarized in Table \ref{tab:fluxes} and can serve as a first approximation for the line intensities and their variability. For the three stronger SO$_2$ lines (258.389 GHz, 258.942 GHz, 259.599 GHz), the intensity is lower in the first observations in December, when compared to the observations in March and April 2017. The NaCl line intensity appears to be more variable on first inspection yet without showing a clear trend. While some of the changes in the intensities derived from the Gaussian fit areas originate from changes in the peak emission, others arise from only changes in the line widths. For a more reliable and comprehensive comparison, we apply an atmospheric model and fit model spectra to the extracted line spectra.

\begin{table}
\caption{Areas under fitted Gaussian profiles for the extracted SO$_2$ and NaCl spectra from the four observing days.}
\begin{tabular}{lcccccc}
\hline
    & \multicolumn{4}{c}{SO$_2$ lines} & & NaCl line \\
Obs     & 258.389 GHz & 258.667 GHz & 258.942 GHz &  259.599 GHz    & &  260.223 GHz	\\
\#  & (K km/s)  & (K km/s)  & (K km/s)  & (K km/s)  & &(K km/s) \\
\hline		
1  &  14.9$\pm$ 1.9 &   6.2$\pm$ 2.1 &  15.7$\pm$ 1.8 &  13.9$\pm$ 1.8 & &  15.2$\pm$ 1.8  \\
2  &  15.3$\pm$ 1.0 &   5.8$\pm$ 0.9 &  18.2$\pm$ 0.8 &  16.7$\pm$ 0.9 & &  12.9$\pm$ 0.9  \\
3  &  18.1$\pm$ 1.5 &   5.9$\pm$ 1.2 &  20.8$\pm$ 1.4 &  17.6$\pm$ 1.2 & &  20.7$\pm$ 1.9  \\
4  &  16.5$\pm$ 1.2 &   5.8$\pm$ 1.3 &  19.4$\pm$ 1.1 &  19.4$\pm$ 1.0 & &  16.1$\pm$ 1.3  \\
\hline
\end{tabular}
\label{tab:fluxes}
\end{table}
  
\section{Atmosphere modelling}
\label{sec:model}

The observed SO$_2$ and NaCl lines are fitted using the radiative transfer model used in \citet{moullet10}. The atmosphere is assumed to be concentrated to a band around the equator with a uniform column density within 35$^\circ$ N/S latitude% and zero density at higher latitudes
, similar to the distribution in \citet{strobel01}. Io's atmosphere is furthermore assumed to be in hydrostatic equilibrium, with a gas temperature horizontally and vertically uniform. Line opacity is calculated for each 0.25~km {\cb thick} layer up to two scale heights, using the transition parameters (intensity and lower level energies) from the JPL and CDMS databases \cite{pickett_jpl,mueller_cdms,endres_cdms_2016} as provided in the Splatalogue catalogue (https://www.cv.nrao.edu/php/splat/). Then local brightness temperature is calculated over a grid of observable disk locations, taking into account airmass, assuming a surface continuum brightness temperature of 100 K \citep{moullet08}. Finally a disk-averaged model line is derived, directly comparable to disk-averaged observations.

{\cb On each observation date, the four simultaneously observed SO$_2$ lines strongly constrain the gas temperature. Specifically, in an optically thin regime, the ratio of line contrasts between two transitions is equal to the ratio of their line intensities. The intensity of a given transition varies with atmospheric temperature, more or less steeply depending on the transition's lower energy level. Hence there is only a limited range of temperature solutions which can reproduce the observed line contrast ratio between two transitions. Thanks to the large span in lower energy levels in the observed lines (from 35 to 360 cm$^{-1}$), the atmospheric temperature can be tightly constrained by finding the best temperature which fits all contrast ratios between different lines. In partially optically thick regimes, one can determine simultaneously the atmospheric temperature and SO$_2$ column density, by considering both the relative and absolute contrasts, with the relative contrasts still being primary diagnostic for the temperature. The line-widths are also sensitive to gas temperature, optical depth (i.e. SO$_2$ column density) and other broadening mechanisms (e.g., planetary-scale winds or plume dynamics).}

{\cb Because the spatial resolution of our observed maps is not sufficient to derive a reliable velocity field, we first assume that the atmosphere is co-rotating with the solid surface (no winds) and retrieve best-fit SO$_2$ column density and temperature.} While the line contrasts could be fit well (except for the weak and noisy 258.667~GHz line on Dec 11), the line-widths were significantly underestimated by our no-wind models. In particular, we noted that in the new IRAM data, almost all SO$_2$ lines are significantly lopsided towards red-shifts. If these red-shift signatures are real, there are different possible explanations for the related global atmospheric motions away from the observer. The dominant red-shift could be related to winds from the observed trailing/anti-Jovian hemisphere towards the opposite (leading/sub-Jovian) hemisphere. Such wind direction could be driven by a pressure gradient from the dayside to the nightside or related to drag forces exerted by the flow of the surrounding plasma from the upstream trailing hemisphere to the downstream leading hemisphere. 

The mismatch in line widths, with broader observed lines than estimated by the models, is a recurring issue in the modeling of Io's atmospheric submillimeter lines. \citet{moullet10} showed that a strong prograde wind, as observed in \cite{moullet08}, could be introduced in the radiative transfer model to fit line-widths. The assumption of prograde winds is not based on a known physical process that could drive such winds, but has a more practical modeling reason. Prograde winds are corotating with Io's rotation, thus producing broader model lines at reasonable wind speeds of a few hundreds of meters per second, similar to the expected sound speed at low altitudes \citep{strobel94}. For a better agreement with the observations, we also included prograde winds in our modeling with the wind speed as free parameter in addition to the gas temperature and column density.
{\cb We emphasize however that our disk-integrated data does not enable a characterization of any wind pattern. Earlier spatially resolved observations by \cite{,moullet08} were not in agreement with sub-solar to anti-solar wind but only with prograde winds. In our observations, a different global velocity field may be at play (especially considering plumes can also have an effect on the line structure. }

The fitted parameters are summarized in Table \ref{tab:results}. Figure \ref{fig:modelSO2} shows the model fits to the four SO$_2$ lines for the four observations, respectively.

\begin{table}
\caption{SO$_2$ and NaCl column densities, temperature and wind parameters derived for the four observation days, assuming an equatorial, homogeneous model atmosphere.}
\centering
\begin{tabular}{lllcccc}
\hline
 Observation&  &  N$_{SO_2}$ &	 T$_{SO_2}$	& v$_{wind}$	&  N$_{NaCl}$	&  Mix. ratio	\\
 Date 			&  &  (cm$^{-2}$)		&   (K)			& (m/s) &	(cm$^{-2}$)	&  NaCl/SO$_2$	\\					
\hline		
%\multicolumn{7}{c}{EXTRA INFO} \\
2016-Dec-11	&	&	$(0.75 \pm 0.11) \times 10^{16}$	& 260	& 250   & 	$(1.40 \pm 0.20) \times 10^{13}$	&	1.9($\pm$0.4)\permil \\
2017-Mar-14	&	&	$(1.10 \pm 0.09) \times 10^{16}$	& 220	& 300	&	$(1.20 \pm 0.09) \times 10^{13}$	&	1.1($\pm$0.1)\permil \\
2017-Apr-06	&	&	$(1.05 \pm 0.12) \times 10^{16}$	& 240	& 250	&	$(1.20 \pm 0.15) \times 10^{13}$	&	1.1($\pm$0.2)\permil \\
2017-Apr-29	&	&	$(1.10 \pm 0.06) \times 10^{16}$    & 255	& 280	&	$(1.30 \pm 0.10) \times 10^{13}$	&	1.1($\pm$0.1)\permil \\
\hline
\end{tabular}
\label{tab:results}
\end{table}

In order to derive NaCl abundance and compare it to the SO$_2$ content, we assume that the two species are colocated and share the same kinetic temperature. However, NaCl is likely highly spatially inhomogeneous and not colocated with SO$_2$ \citep{moullet15}, since SO$_2$ is mainly sourced from sublimation while NaCl originates from volcanic activity. The assumption of a global abundance leads to lower column densities {\cb and the obtained NaCl values in Table \ref{tab:results} can in that sense be considered lower limits.}

The other extreme case would be high NaCl column densities in only few small regions on the surface (at active volcanoes), which means higher optical depth and at some NaCl density saturated lines. However, we have no information on the distribution or degree of confinement of the NaCl abundance. The goal of the study is to constrain the temporal evolution of the atmosphere. Therefore, we chose to neglect spatial variations and used the simplest assumption, keeping the shortcomings in mind. 

The model fits to the NaCl lines for the four observations are shown in the rightmost column of Figure \ref{fig:modelSO2}.

\section{Results and discussion}
\label{sec:discuss}

\subsection{Atmosphere}

The obtained SO$_2$ temperatures and abundances are in general agreement with previous submillimeter and Lyman-$\alpha$ observations of the trailing hemisphere \citep[e.g.][]{moullet08,feaga09,moullet10}. {\cb We note that there are however some difference in obtained temperature and density when comparing to values using other methods. For example, higher densities are often derived from thermal infrared and near-ultraviolet measurements \citep[e.g.,][]{spencer05,tsang12,jessupspencer15} and solar reflected mid-infrared spectra suggest lower temperatures \citep{lellouch15}. Such differences can be caused by differing sounding altitudes in the atmosphere for example.}

During the three observations taken in March and April 2017 the SO$_2$ column density appears to be stable, i.e. the results are consistent with a constant abundance with the 1-$\sigma$ uncertainties (Figure \ref{fig:compare}c and Table \ref{tab:results}). Compared to the average of the three 2017 observations, the  SO$_2$ column density derived for December 2016 is lower by 30($\pm14$)\%, suggesting a significant ($\sim$2$\sigma$) change. Comparing the combined fluxes directly derived from the four SO$_2$ spectra (Table \ref{tab:fluxes}), the Dec 2016 value is also lower than the average from the March and April 2017 tracks (by  15($\pm7$)\% as compared to 30($\pm14$)\% from the model), suggesting that a change in the SO$_2$ atmosphere did happen in this period.

In contrast to this trend, the highest NaCl column density was derived for the December 2016 observation (Table \ref{tab:results}). However, the modeled NaCl abundance appears overall to be stable and all four observations are consistent with the mean value within their 1-$\sigma$ uncertainties (green dotted line in Figure \ref{fig:compare}c). Given the discussed short NaCl lifetime \citep{moses02-2} and likely dynamic volcanic sources, the stability is generally surprising.  Again looking at fluxes under the NaCl spectra (Table \ref{tab:fluxes}), they undergo more variability with a peak on April 6 between the March 14 and April 29 observations. The maximum originates from the wide red-shifted wing which is not captured by the model (Figure \ref{fig:modelSO2}, right column). If real, this extended wing might originate from a specific eruption, e.g. located on the dusk hemisphere which moves away from the observer. 

The resulting NaCl/SO$_2$ mixing ratio are on the low side of previously derived ratios \citep{lellouch03,moullet10}. As discussed in \cite{lellouch03} the derived values can vary by more than an order of magnitude for different assumptions on the NaCl distribution. When assuming a localized abundance only in plumes the fitted NaCl abundance would be higher. 

Given the trends in SO$_2$ and NaCl, the NaCl/SO$_2$ mixing ratio decreases from 11 December 2016 to the observations taken on 14 March, 6 and 29 April in 2017 (Table \ref{tab:results}, rightmost column). During the period in March and April it remains stable within uncertainties. Again relating the difference between December 2016 and March/April 2017 to the stable 2017 ratio, we find a relative decrease by 64\%($\pm$36\%).

The stability of the NaCl column density might be interpreted as a quiet atmospheric state, where volcanic outgassing sources are small. Low volcanic activity would imply low supply of NaCl through outgassing. The increase in SO$_2$ abundance on the contrary suggests an increase in the atmospheric sources related to volcanic activity. Seasonal changes (due to the heliocentric distance) and related changes in sublimation yields are negligible within the studied period.

In the following sections, we compare our results with the variability observed in the presence of hot spots on Io, the Jovian neutral sodium cloud, the sulfur ion torus and Jupiter's polar aurora. 

\subsection{Hot spot activity}

\begin{figure}
\center\noindent\includegraphics[width=1.\textwidth]{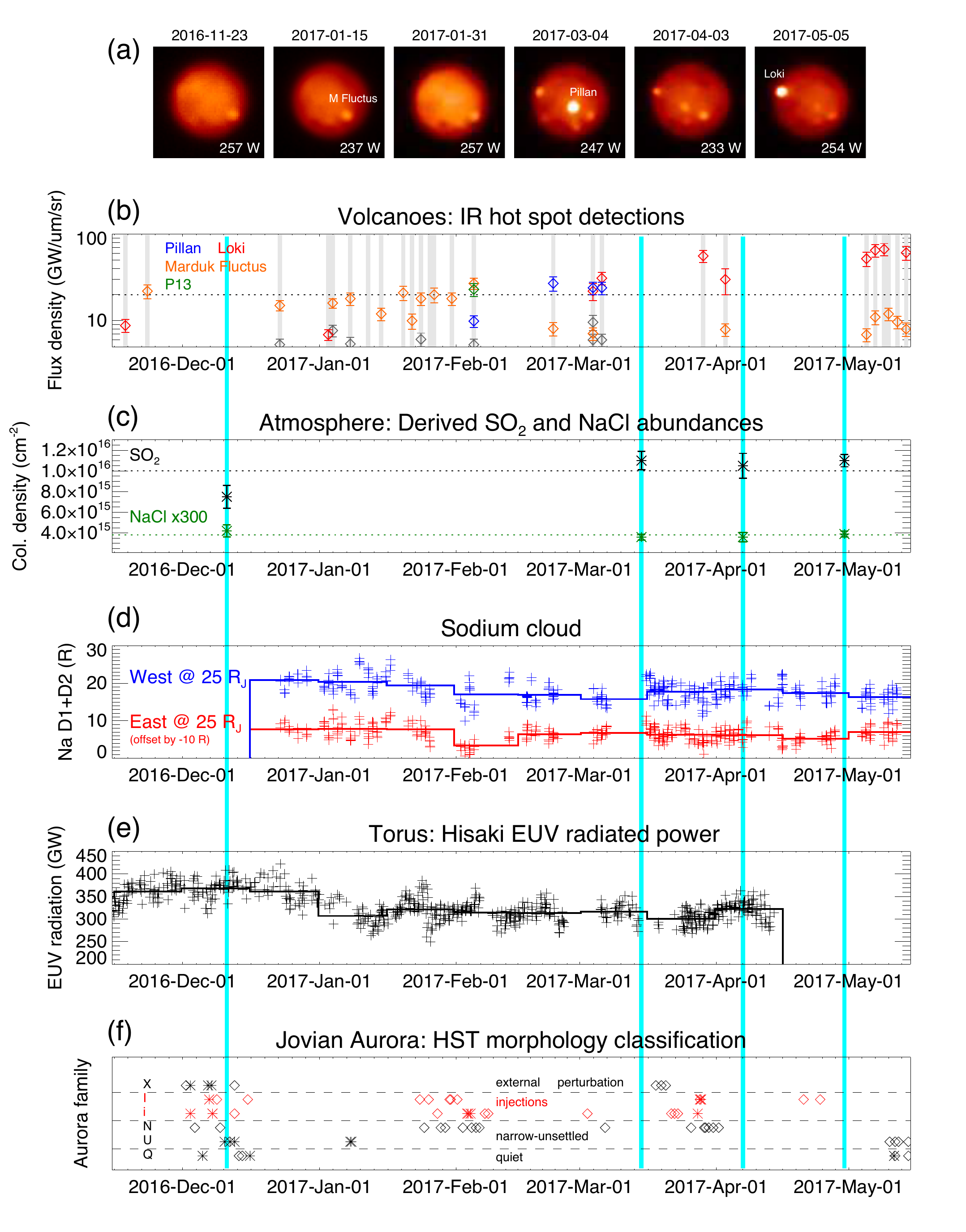}
\caption{Comparison of various measurements obtained in the period from 15 Nov 2016 until 15 May 2017 that are possibly related to Io's atmosphere: (a) Gemini images {\cb (3.8 $\mu$m)} of volcanic hot spots on the hemisphere (CML in white) observed by IRAM (Table \ref{tab:obs}) (b) Hot spot brightnesses extracted for specific volcanic regions (colored) from all Gemini and Keck IR images that covered the studied hemisphere (dates of used images shown by grey vertical line).  (c) Atmospheric column densities for SO$_2$ (black) and NaCl (green, multiplied with 300) on the four days of our IRAM observations and averages (dotted lines). (d) Brightness of the Na D$_1$ and D$_2$ lines of the sodium cloud measured east (red crosses) and west (blue crosses) of Jupiter and half-month averages (solid lines).  (e) Total radiated power of the Io torus (dominated by sulfur ion emissions) measured in the EUV channel of the Hisaki space observatory (crosses) and half-month averages (solid line). (f) Classification timeline of the morphology of Jupiter's aurora from HST images by \cite{grodent18}.  The "injection" classes (i,I, in red) are suggested to be related to enhanced mass loss from Io's atmosphere. The light blue vertical bars are added to guide the eye for the comparison with the IRAM observations.}
\label{fig:compare}
\end{figure}

The thermal emission from volcanic hot spots has been monitored since 2013 with high cadence by the Keck and Gemini telescopes \citep{dekleer16a,dekleer19-iotime}. Here, we focus on the brighter hot spots on the hemisphere covered by IRAM and use the power measured in the L' filter (3.8 $\mu$m) as diagnostic for the activity level. In 26 IR images taken between mid-November 2016 and mid-May 2017 the hemisphere targeted by IRAM was (partly) observed by Gemini or Keck. The times of these 26 observations are shown by grey (sometimes overlapping) vertical lines in Figure \ref{fig:compare}b \citep[for complete list of images and CMLs see table 4 in][]{dekleer19-iotime}. A selection of six Gemini images is shown in the top panel (a). Hot spot emission in the category of bright eruptions ($>$20 GW/$\mu$m/sr) as defined in \cite{dekleer19-iotime} were measured on four different locations: Marduk Fluctus (orange diamonds in Figure \ref{fig:compare}b), Pillan Patera (blue), Loki Patera (red), and an unnamed patera designated P13 (green). The power of all other hot spots (grey) did not exceed 20 GW/$\mu$m/sr, even when including the measurements of the opposite hemisphere not shown in the figure. We first discuss possible relations to the changes at Loki Patera, because of its particular role as periodically {\cb brightening patera} \citep{dekleer19-loki} and because it is the brightest hot spot in this period. 

\subsubsection{Loki Patera}
\cite{mendillo04} suggested that the activity level at Loki is positively correlated to the Jovian sodium cloud brightness through the volcanic outgassing of NaCl to the atmosphere. In our studied period, Loki awakened after a quiet period reaching strong emissions near the later IRAM observations in April. The NaCl abundance is stable and the NaCl/SO$_2$ mixing ratio even significantly drops in the period of the increased thermal activity at Loki. Thus, our results dispute the hypothesis that Loki feeds the sodium cloud through outgassing of NaCl.

On the other hand, the increasing activity at Loki Patera coincides with the obtained increase in SO$_2$ abundance, possibly suggesting a positive correlation of volatile (SO$_2$) abundance and thermal hot spot brightness. However, this is in fact rather unexpected, based on the current understanding of the periodic brightening at Loki Patera as overturning lava lake \citep{rathbun02,davies03, dekleer17}. The volatile content in the repeatedly erupting lava is instead expected to be low and thus thermal eruptions not be accompanied by outgassing events. The lack of spatial resolution in the IRAM observations, however, prevents further investigations of the correlation of the minor increase in SO$_2$ and the wakening of the Loki Patera. 

\subsubsection{Other volcanic spots}

Apart from Loki, activity via hot spots detections was found at Pillan patera and the patera designated P13. The bright eruption at P13 was seen only in one image (February 5) and the power had already significantly decreased until February 23, where 2.2 GW/$\mu$m/sr (L'-band) were measured. It is unlikely that this activity affected the atmosphere density measured on March 14 or later by IRAM, given the atmospheric SO$_2$ lifetime of a few days \citep{strobel01,lellouch07}.

The transient brightening at the Pillan Patera was seen in late February and early March. A faint signal from the site can be seen in the Gemini image from 3 April (Figure \ref{fig:compare}a). Hence, the temporal coincidence makes Pillan a possible candidate for causing the increase in SO$_2$ abundance between Dec 2016 and March 14. However, previous observations did not reveal a measurable effect on the SO$_2$ abundance during an even brighter eruption at Pillan patera \citep{lellouch15}. The IR images from early March revealed a few other hot spots with L'-band brightnesses $>$5 GW/$\mu$m/sr (Figure \ref{fig:compare}b, grey diamonds), which potentially could be related to the SO$_2$ increase. 

The Marduk Fluctus hot spot on the other hand remains relatively bright over the observed period, similar to the behavior observed in the years before the studied period \citep{dekleer16a}. A measurable change of the SO$_2$ abundance due to the activity is thus not expected.

We note also that the maximum power of the bright hot spots (other than Loki) did not reach the high level of other events detected in the years before and after \citep{dekleer14,dekleer19-iotime}. 

{\cb Taken together, the presented data sets do not allow a reliable conclusion on the relation between hot spots and atmospheric abundances. The only detectable change in the IRAM data occurred during the long observational gap between Dec 2016 and March 2017 and it is generally difficult to relate this change to some volcanic event in this longer period. Given the short atmospheric life times, a volcanic outburst might lead to only transient changes over a few days. The sparsity of hot spot observations directly before the IRAM observations in March and April 2017 further hampered the interpretation.}

\subsection{Magnetospheric neutral and plasma environment}

\subsubsection{Sodium cloud}

The neutral sodium cloud across Jupiter's magnetosphere was observed frequently between 23 Dec 2016 until later in 2017 from Mt. Haleakal$\bar{\mathrm a}$, Maui, Hawaii.  The measured brightness relates to the resonantly scattered sunlight from the sodium D$_1$ and D$_2$ lines. For more details on instrument, method and data processing, see \citep{yoneda09}. Figure \ref{fig:compare}d shows each measurement on the eastern (red crosses) and western sides (blue crosses) of Jupiter at a distance of 25 R$_J$ to the planet. The majority of the measurements are consistent (within the measurement uncertainties, not shown in plot) with the average brightnesses of 17 R (East) and 18 R (West) of this period. These average brightnesses are also similar to values previously found for periods of stable sodium cloud brightness \cite{yoneda09}. 

When comparing the early period (23 Dec to 15 Jan) to the period of March and April 2017, there is a minor decrease of 10\% (East) and 14\% (West). This could be related to the decrease in NaCl/SO$_2$ mixing ratio in the IRAM data (Table \ref{tab:results}). However, a change in total abundance of NaCl in the atmosphere was not derived and the only minor change in the cloud prevent further conclusions.  For comparison, in an extreme case in early 2015, these sodium brightnesses increased to values higher than 60 R within about one month \citep{yoneda15}. Therefore, the period studied here shows a quiescent sodium cloud. 

Of the detected bright volcanic eruptions, none appears to have affected the sodium cloud. In the same time period where the sodium cloud slightly decreases, the volcanic activity overall rather undergoes an increase (see discussion above and panels a and b in Figure \ref{fig:compare}). In particularly, the Loki Patera, where the activity strongly increased from February 2017 (hardly detectable) until May 2017 ($>$80 GW/$\mu$m/sr), seems to be uncorrelated to the sodium cloud.

\subsubsection{Plasma and neutral torus}

Next, we look at the sulfur ion emissions from the Io torus, which was monitored by the Hisaki satellite \citep{yoshikawa14} from November 2016 until 13 April 2017. Figure \ref{fig:compare}e shows the total radiation power of the Io plasma torus integrated over the wavelengths between 65 and 78 nm, including line emissions of sulfur ions, S$^+$, S$^{2+}$, and S$^{3+}$ \citep{kimura17}. We used the level-2 spectrograph image, for which detected photons are accumulated for 1 minute \citep{kimura19}. The level-2 images obtained during one Hisaki orbital period around the Earth($\sim$106 min) were integrated with resulting total integration times of 30-60 minutes.

There is a decrease by 15($\pm$8)\% from the average brightness of 370 GW in December 2016 down to 310 GW averaged over March and April 2017. Even though this trend is clear and significant, it is again only a minor change when compared for example to the event in 2015 where the power almost doubled in less than two months \citep[e.g.,][]{yoshioka18,kimura18}.

The decrease in the torus coincides with the observed 30\% increase in SO$_2$ abundance in the atmosphere. The time scales derived for the changes in the neutral and plasma {\cb torus emissions measured by Hisaki in 2015} are between 10 and 40 days \citep{yoshioka18,koga19}. The different trends can thus hardly be explained by temporal delay of the response from the torus. The anti-correlation disagrees with the general understanding that higher atmospheric abundances leads to higher mass loss and thus higher torus. {\cb The torus trend could originate e.g. from changes in the electron temperature which affects the torus EUV emission in addition to the torus densities. In optical torus emission measurements, which are more sensitive to density, \cite{schmidt18} did not report unusual SII brightness in the priod March-May 2017 either.}

The torus brightness decrease is, however, similar to the found decrease in the sodium cloud brightness (although the sodium monitoring started only later on Dec 23). {\cb This similar trend in ion torus emissions and sodium emissions together with opposed trend in the SO$_2$ atmosphere is more consistent with a change that originated in the magnetosphere but inconsistent with an atmospheric change as trigger.}

The Hisaki data on the neutral oxygen torus emissions \citep{koga18a,koga18b} from the studied period are not yet calibrated and are therefore not included in this discussion.

\subsubsection{Aurora activity}

Using the Hubble Space Telescope (HST), \cite{grodent18} systematically imaged Jupiter's aurora in the period between 30 November 2016 and 18 July 2017. They systematically classified the observed emission morphologies into six aurora families.  Figure \ref{fig:compare}f shows the changes between the morphology families 'quiet' (Q), unsettled (U), narrow (N), injection (i, in orange), strong injection (I, in red), and external perturbation (X). The two 'injection' morphologies were interpreted to arise from magnetospheric plasma injections into the magnetosphere from Io after a strong volcanic eruption (or several strong eruptions) \citep{grodent18,bonfond12}. 

HST detected clustered 'injections' morphologies (including 'strong injection') during several periods in late 2016 and early 2017. One such period happened around the first IRAM observation in December 2016, where the SO$_2$ abundance was lowest. Another one between the second and third IRAM observation. None of these periods with multiple 'injection' morphology detections seem to be reflected by any of the other data discussed here, although the observing times are not always overlapping. For example, there were frequent IR observations of Io's volcanic activity before the 'strong injection' aurora period in late January 2017, but the only brighter detected hot spot was at the 'Marduk Fluctus' region, which had been constantly active long before (see panel b). Further comparisons are difficult, as the aurora morphology is very dynamic and changes quickly on time scales of days. These changes are probably dominated by various processes of the magnetospheric dynamics and not by the mass loss from Io. 

An aurora morphology similar to the 'injection' class that did not coincide with any known volcanic activity of magnetospheric changes was observed before for example by \cite{badman16}. Explanations other than enhanced mass loading for the particular aurora morphology could be that a global magnetospheric reconfiguration that includes the inward injections is triggered further out in the magnetosphere \citep{louarn14, haggerty19}.

\section{Summary and open questions}
\label{sec:sum}

We have observed line emissions from SO$_2$ and NaCl in Io's atmosphere over roughly the same hemisphere on four days between 11 December 2016 and 29 April 2017. By fitting simulated fluxes from an atmosphere model to the observed fluxes, we find that the SO$_2$ abundance on the three days in 2017 is stable with a column density of N$_{SO_2} = 1.1 \times 10^{16}$ cm$^{-2}$. Compared to this stable period, a 30\% lower column density was derived for December 2016. The model fits for the simultaneously measured NaCl line revealed a roughly stable NaCl abundance of N$_{NaCl} = 1.2-1.4 \times 10^{13}$ cm$^{-2}$ for all four dates. Our abundances are generally consistent with earlier observations, but we note that the absolute abundances of the species depend on model assumptions, and the focus of this work is on relative changes between the observations. 

The increase in SO$_2$ from December 2016 to March/April 2017 can not be explained by seasonal variations, suggesting that it is instead related to changes in the volcanic source of the atmosphere. The constant abundance of the volcanic trace gas NaCl (and thus decreasing NaCl/SO$_2$ mixing ratio) does however not support this possible change in volcanic supply to the atmosphere. Although the change in SO$_2$ has a 2-$\sigma$ significance, we note that the observing conditions were least favorable during the Dec 2016 track (unfavorable weather, smallest angular size of Io) possibly affecting the signal.    

The change in the bulk SO$_2$ atmosphere observed by IRAM coincides with a significant increase of the Loki hot spot brightness. However, a significant effect on the SO$_2$ from Loki is not expected given our understanding of the patera as periodically overturning lava lake \citet{rathbun02}. 

\cite{mendillo04} had suggested that the Jovian sodium cloud brightness is correlated to Loki's activity via the outgassing of NaCl. Dissociation of NaCl in Io's atmosphere and production of fast Na atoms is generally agreed on to be the primary source for the sodium cloud. A change in the sodium cloud would thus require a change in atmospheric NaCl, which is not seen in the IRAM data. {\cb Furthermore, the sodium cloud brightness became fainter during the period of the wakening of the activity at Loki and the EUV torus brightnesses faded as well. Taken together, the presented data sets contradict the results of the study by \cite{mendillo04}, which is often cited as main evidence for a connection of volcanic activity to mass loss from the moon.}

The other hot spot activity during the observed period revealed considerable brightenings (at the lower end of the "bright eruption" category of \cite{dekleer19-iotime}) at two sites (Pillan Patera and P13) between the first and later IRAM observations. The eruptions at these hot spots could possibly be related to the increase in the bulk SO$_2$ atmosphere, but the limited coverage and overlap prevent further conclusions.

The monitoring of the Jovian sodium cloud and Io plasma torus in general revealed a quiescent {\cb (i.e., no transient changes)} environment, consistent with a stable atmosphere. Observed frequent appearances of a fainter, more extended main emissions in Jupiter's aurora, earlier suggested to relate to increased mass loading {\cb \citep{bonfond12}}, are not reflected in any of the other data and are thus likely unrelated. 

Taken together, both the atmosphere and the Jovian environment can be considered quiescent in the studied period. This prevents further conclusions on the volcanic changes of the magnetospheric environment. Hence, the detection of an unambiguous transient volcanic change in Io's bulk atmosphere remains an unresolved task. Independently of the observational evidence, key questions on the influence of changes in Io's volcanic activity on the magnetospheric environment through the atmospheric mass loss remain unanswered:
\begin{itemize}
\item Can changes in volcanic activity lead to significant changes in the bulk atmospheric loss?
\item If so, what are the characteristics of such 'volcanic mass loading events' and under what conditions do they happen? 
\item What are the time scales of the involved processes? 
\end{itemize}
Possible scenarios could be that the outgassing level from the volcanic sites undergoes global changes that are not directly observable through hot spot emissions or otherwise. 
The location of the outgassing sites could also play a role for the effectiveness of atmospheric sputtering to allow the plume gas to escape. However, a change in mass loading from Io's atmosphere through the ion-neutral collisions can also be caused by an 'external trigger' in plasma torus properties like an increase in density not triggered or initiated by Io's volcanic activity (as mentioned in the introduction).
In order to address and possibly resolve this issue in the future, continuous dedicated monitoring of key atmospheric species (such as SO$_2$ and NaCl) with spatial resolution across Io's disk and optimally during a period of strong changes in the plasma environment is needed.

\section*{Acknowledgements}

L.~R. appreciates the support from the Swedish National Space Agency (SNSA) through grant 154/17 and the Swedish Research Council (VR) through grant 2017-04897. A.~S.-M., P.~S. and S.~T. (Cologne) have been supported via Collaborative Research Centre 956, funded by the Deutsche Forschungsgemeinschaft (DFG; project ID 184018867) and DFG SCHL 341/15-1 (“Cologne Center for Terahertz Spectroscopy”).

\section*{References}

% Journal abbreviation definitions
\def\aj{AJ}%
          % Astronomical Journal
\def\actaa{Acta Astron.}%
          % Acta Astronomica
\def\araa{ARA\&A}%
          % Annual Review of Astron and Astrophys
\def\apj{ApJ}%
          % Astrophysical Journal
\def\apjl{ApJ}%
          % Astrophysical Journal, Letters
\def\apjs{ApJS}%
          % Astrophysical Journal, Supplement
\def\ao{Appl.~Opt.}%
          % Applied Optics
\def\apss{Ap\&SS}%
          % Astrophysics and Space Science
\def\aap{A\&A}%
          % Astronomy and Astrophysics
\def\aapr{A\&A~Rev.}%
          % Astronomy and Astrophysics Reviews
\def\aaps{A\&AS}%
          % Astronomy and Astrophysics, Supplement
\def\azh{AZh}%
          % Astronomicheskii Zhurnal
\def\baas{BAAS}%
          % Bulletin of the AAS
\def\bac{Bull. astr. Inst. Czechosl.}%
          % Bulletin of the Astronomical Institutes of Czechoslovakia 
\def\caa{Chinese Astron. Astrophys.}%
          % Chinese Astronomy and Astrophysics
\def\cjaa{Chinese J. Astron. Astrophys.}%
          % Chinese Journal of Astronomy and Astrophysics
\def\icarus{Icarus}%
          % Icarus
\def\jcap{J. Cosmology Astropart. Phys.}%
          % Journal of Cosmology and Astroparticle Physics
\def\jrasc{JRASC}%
          % Journal of the RAS of Canada
\def\mnras{MNRAS}%
          % Monthly Notices of the RAS
\def\memras{MmRAS}%
          % Memoirs of the RAS
\def\na{New A}%
          % New Astronomy
\def\nar{New A Rev.}%
          % New Astronomy Review
\def\pasa{PASA}%
          % Publications of the Astron. Soc. of Australia
\def\pra{Phys.~Rev.~A}%
          % Physical Review A: General Physics
\def\prb{Phys.~Rev.~B}%
          % Physical Review B: Solid State
\def\prc{Phys.~Rev.~C}%
          % Physical Review C
\def\prd{Phys.~Rev.~D}%
          % Physical Review D
\def\pre{Phys.~Rev.~E}%
          % Physical Review E
\def\prl{Phys.~Rev.~Lett.}%
          % Physical Review Letters
\def\pasp{PASP}%
          % Publications of the ASP
\def\pasj{PASJ}%
          % Publications of the ASJ
\def\qjras{QJRAS}%
          % Quarterly Journal of the RAS
\def\rmxaa{Rev. Mexicana Astron. Astrofis.}%
          % Revista Mexicana de Astronomia y Astrofisica
\def\skytel{S\&T}%
          % Sky and Telescope
\def\solphys{Sol.~Phys.}%
          % Solar Physics
\def\sovast{Soviet~Ast.}%
          % Soviet Astronomy
\def\ssr{Space~Sci.~Rev.}%
          % Space Science Reviews
\def\zap{ZAp}%
          % Zeitschrift fuer Astrophysik
\def\nat{Nature}%
          % Nature
\def\iaucirc{IAU~Circ.}%
          % IAU Cirulars
\def\aplett{Astrophys.~Lett.}%
          % Astrophysics Letters
\def\apspr{Astrophys.~Space~Phys.~Res.}%
          % Astrophysics Space Physics Research
\def\bain{Bull.~Astron.~Inst.~Netherlands}%
          % Bulletin Astronomical Institute of the Netherlands
\def\fcp{Fund.~Cosmic~Phys.}%
          % Fundamental Cosmic Physics
\def\gca{Geochim.~Cosmochim.~Acta}%
          % Geochimica Cosmochimica Acta
\def\grl{Geophys.~Res.~Lett.}%
          % Geophysics Research Letters
\def\jcp{J.~Chem.~Phys.}%
          % Journal of Chemical Physics
\def\jgr{J.~Geophys.~Res.}%
          % Journal of Geophysics Research
\def\jqsrt{J.~Quant.~Spec.~Radiat.~Transf.}%
          % Journal of Quantitiative Spectroscopy and Radiative Trasfer
\def\memsai{Mem.~Soc.~Astron.~Italiana}%
          % Mem. Societa Astronomica Italiana
\def\nphysa{Nucl.~Phys.~A}%
          % Nuclear Physics A
\def\physrep{Phys.~Rep.}%
          % Physics Reports
\def\physscr{Phys.~Scr}%
          % Physica Scripta
\def\planss{Planet.~Space~Sci.}%
          % Planetary Space Science
\def\procspie{Proc.~SPIE}%
          % Proceedings of the SPIE
\let\astap=\aap
\let\apjlett=\apjl
\let\apjsupp=\apjs
\let\applopt=\ao
%\bibliographystyle{/Users/lorenzroth/Dropbox/Work/Bib/agufull08}
%\bibliography{bib2019}

\end{document}